\documentclass[onecolumn,12pt,conference]{IEEEtran}
\usepackage{longtable}
\usepackage{amsmath}
\usepackage{cancel}
\usepackage{amssymb}
\usepackage{graphicx}
\usepackage{longtable}
\usepackage{multirow}
\usepackage{array}
\usepackage[labelsep=period]{caption}
\usepackage{setspace}
\usepackage{float}
\usepackage{subfig}
\usepackage{lettrine}
\usepackage[noadjust]{cite}
\usepackage{cleveref}
\newcommand{\qed}{\nobreak \ifvmode \relax \else
	\ifdim\lastskip<1.5em \hskip-\lastskip
	\hskip1.5em plus0em minus0.5em \fi \nobreak
	\vrule height0.75em width0.5em depth0.25em\fi}

\begin{document}
\title{On the Capacity of Wireless Powered Cognitive Relay Network with Interference Alignment}
\author{$\textrm{Sultangali~Arzykulov}^{*}$,~ $\textrm{Galymzhan~Nauryzbayev}$,~$\textrm{Theodoros A. Tsiftsis}^{*}$ \\
	and
$\textrm{Mohamed~Abdallah}^{\ddagger}$\\
	\IEEEauthorblockA{$^*$School of Engineering, Nazarbayev University, Astana, Kazakhstan\\
	$^{\ddagger}$Division of Information and Computing Technology, College of Science and Engineering, \\
	Hamad Bin Khalifa University, Qatar Foundation, Doha, Qatar\\
	Email: \{sultangali.arzykulov, theodoros.tsiftsis\}@nu.edu.kz,~nauryzbayevg@gmail.com,\\
	moabdallah@hbku.edu.qa}
}
\maketitle

\begin{abstract}
In this paper, a two-hop decode-and-forward cognitive radio system with deployed interference alignment is considered. The relay node is energy-constrained and scavenges the energy from the interference signals. In the literature, there are two main energy harvesting protocols, namely, time-switching relaying and power-splitting relaying. We first demonstrate how to design the beamforming matrices for the considered primary and secondary networks. Then, the system capacity under perfect and imperfect channel state information scenarios, considering different portions of time-switching and power-splitting protocols, is estimated. 
\end{abstract}

\IEEEpeerreviewmaketitle

\section{Introduction}
\lettrine{C}{ognitive} radio (CR) is a promising paradigm that can resolve the problem of the spectrum scarcity \cite{Haykin}. Cognitive radio network (CRN) consists of primary users (PUs), who are licensed users of the spectrum, and secondary users (SUs), who can access the spectrum only when they do not cause harmful interference to PUs \cite{Goldsmith}. SUs can access the licensed spectrum by three paradigms such as  interweave, underlay and overlay \cite{Goldsmith}. Interference alignment (IA) is a very powerful technique potential to provide higher degrees of freedom (DoFs) which means interference-free signaling dimensions at receivers \cite{galym1,galym2,GS1, GS2,Cadambe}. The promising performance of IA can be also attained in CR by suppressing interference at PUs. By using this technique, SUs easily coexist with primary networks (PNs) by achieving higher data rates and not causing harmful interference to PUs \cite{Amir}. Many previous studies introduced the impact of IA on the performance of the underlay CR \cite{Yi,Xu,Zhao}. A multiple-input multiple-output (MIMO) CRN with cooperative relay was presented in \cite{Tang1}, where DoFs for the network was increased by using IA. The authors in \cite{Chen} and \cite{Ding} introduced energy-harvesting (EH) as a promising technology to prolong the life-time of battery-powered wireless devices. There are two main EH architectures such as time-switching (TS) and power-splitting (PS) protocols \cite{Nasir}. In \cite{Zhao1}, the authors proposed a common framework to jointly study IA and simultaneous wireless information and power transfer (SWIPT) in MIMO networks, where the users were dynamically chosen as EH terminals to improve the performance of wireless power transfer (WPT) and information transmission. The work in \cite{Park} considered a spectrum sensing policy for the EH-based CR to ensure that SUs can harvest the energy from the PUs signals in the TS mode. In \cite{Zheng}, the same system was investigated with the optimal information and energy cooperation methods between SUs and PUs, where SUs harvest energy from PUs, then use that energy to transmit their own and PUs' signals. Both \cite{Park} and \cite{Zheng} assumed that all nodes in CR have a perfect channel state information (CSI), however, in practice it is more common to deal with imperfect CSI due to channel estimation errors \cite{Wang}.

Unlike the existing studies and to the best of our knowledge, no work has been addressed to jointly study these three important areas, namely, CRN, WPT and IA. Therefore, this work focuses on studying the underlay IA-based CRN where an energy-restricted relay node operates in the time-switching relaying (TSR) and power-splitting relaying (PSR) modes. The performance of primary receivers and the EH-based relay of the secondary network (SN) under interference environment is analyzed. Interference at the PUs and the relay is cancelled by the well-known IA technique. The network capacity for these protocols is evaluated with different settings of TS/PS portions and under various CSI scenarios. 


\section{System Model}
\label{sec:system model}

\begin{figure}[!t]
	\centering
	\includegraphics[width=0.4\columnwidth]{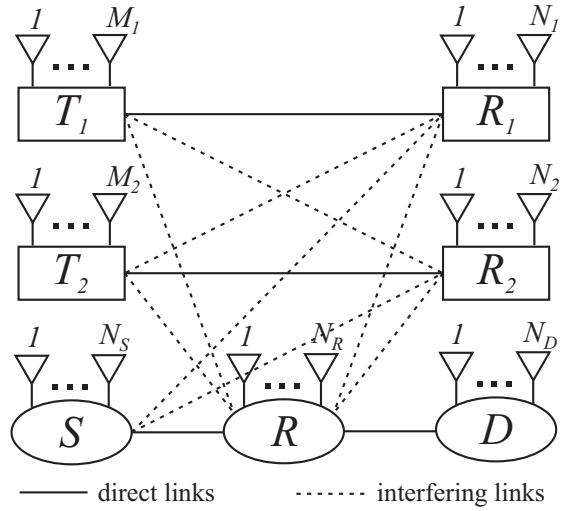}
	\caption{The IA- and EH-based CRN with two PUs and one SU sharing the spectrum simultaneously.}
	\label{system_model}
\end{figure}
We consider a system model consisting of two pairs of PUs and three SUs. Within the PN, each transmitter communicates to its corresponding receiver and causes interference to another primary receiver and secondary relay node. It is assumed that there is no temperature constraint at the SUs which may imply interference at the primary receivers. The SN consists of a source ($S$), a relay ($R$) and a destination ($D$) nodes. $S$ communicates with $D$ through the assistance of the energy-constrained $R$ that operates in a half-duplex mode and decodes and forwards (DF) the signal of $S$ to $D$ within two time slots. To support this, $R$ harvests energy from the interference signals, while $S$ and $D$ are assumed to have external power sources. Furthermore, D is assumed to be located far from PUs and does not receive any interference. Each node of the network is assumed to be deployed with multiple antennas as shown in Fig.\ref{system_model}, where solid lines indicate the direct channels while the dotted lines denote the interfering links. We also assume that the channels remain constant during a transmission block time $T$, but vary independently from one block to another. Channel links between nodes are defined as follows. For the PN case, $\mathbf{H}_{j,i}^{[k]}\in\mathbb{C}^{N_j\times M_i},~\forall i,j \in \{1,2\}$ denotes the channel between receiver $j$th ($R_j$) and transmitter $i$th ($T_i$), where superscript $k$ indicates a certain time period (TP)  when the data transmission occurs. $N_j$ and $M_i$ are the numbers of antennas at $R_j$ and $T_i$, respectively. For the SN case, $\mathbf{H}_{R,S}$ and $\mathbf{H}_{D,R}$ denote the channel links related to the $S$-$R$ and $R$-$D$ transmissions while the inter-network channels are given by $\mathbf{H}_{j,R}\in\mathbb{C}^{N_j\times N_R}$, $\mathbf{H}_{j,S}\in\mathbb{C}^{N_j\times N_S}$ and $\mathbf{H}_{R,i}\in\mathbb{C}^{N_R\times M_i}$, where $N_S$, $N_R$ and $N_D$ denote the numbers of antennas at $S$, $R$ and $D$, respectively. Each entry of any matrix $\mathbf{H}$ is assumed to be independent and identically distributed (i.i.d.) random variables according to $\mathcal{CN}(0,1)$, where $\mathcal{CN}(0,1)$ denotes the complex normal distribution with zero mean and unit variance. Also, note that each channel link is characterized by the corresponding distance and path-loss exponent denoted by $d_{m,n}$ and $\tau_{m,n},~\forall m\in\{1,2,R,D\},~\forall n\in\mathcal{A} = \{1,2,S,R\}$, respectively. Moreover, a global CSI is also assumed.  

We assume that the SN and PN nodes exploit the IA technique to mitigate the interference at $R$ and $R_{i}$, respectively. Thus, any transmit node $l$ with power $P_l$ utilizes a transmit BF matrix $\mathbf{V}_{l}\in\mathbb{C}^{(M_l~\textrm{or}~N_l)\times f_l}$, with  $\textrm{trace}\{\mathbf{V}_l\mathbf{V}_l^H\}=1,~\forall l \in \mathcal{A}$, where $f_l$ is the number of the transmitted data streams. At the same time, all receive nodes (except $D$) exploit the receive BF matrix $\mathbf{U}_t \in\mathbb{C}^{N_t\times f_t},~\forall t\in\{1,2,R\}$, where $f_t$ is the number of data streams to be decoded at the corresponding receiver. For simplicity, we assume that each node is deployed with $N$ antennas ($M_i=N_j=N_{S}=N_{R}=N_{D} = N$). 

The primary receivers,  within two TPs,  obtain the following signal
\begin{multline} 
\label{y_j}
\mathbf{y}_{j}^{[k]} =\underbrace{\sqrt{\frac{P_{j}}{d_{j,j}^{\tau_{j,j}}}} \mathbf{U}^{[k]H}_{j} \mathbf{H}^{[k]}_{j,j}\mathbf{V}^{[k]}_{j}\mathbf{s}_{j}}_{\text{desired signal}} + \underbrace{\mathbf{T}^{[k]}}_{\text{interference from SN}}
 + \underbrace{ \sqrt{\frac{P_{i}}{d_{j,i}^{\tau_{j,i}}}} \mathbf{U}^{[k]H}_{j}\mathbf{H}^{[k]}_{j,i}\mathbf{V}^{[k]}_{i}\mathbf{s}_{i}}_{\text{interference from PN, \( i\neq j \)}}+ {\tilde{\mathbf{n}}}^{[k]}_{j},~k\in\{1,2\},
\end{multline}
where $\tilde{\mathbf{n}}^{[k]}_{j}=\mathbf{U}^{[k]\,H}_{j}\mathbf{n}^{[k]}_{j}$ is the effective zero-mean additive white Gaussian noise (AWGN) vector at the output of the beamformer, with $\mathbb{E}\{\tilde{\mathbf{n}}^{[k]}_{j}\tilde{\mathbf{n}}^{[k]H}_{j}\}=\sigma^2_{\tilde{n}_j}\mathbf{I}$, where $\mathbb{E}\{\cdot\}$ denotes an expectation operator. Since we assume that $\mathbf{s}_l$ is the vector containing the symbols drawn from i.i.d. Gaussian input signalling and chosen from a desired constellation, we have $\mathbb{E}\{\mathbf{s}_l\mathbf{s}_l^H\}=\mathbf{I}$, with $l\in\mathcal{A}$. Meeting all these conditions satisfies the average power constraint at each transmit node. Regarding the SN, its interference to PN can be defined as
\begin{equation}
\mathbf{T}^{[k]} = \begin{cases}
 \sqrt{\frac{P_{S}}{d_{j,S}^{\tau_{j,S}}}} \mathbf{U}^{[k]H}_{j}\mathbf{H}_{j,S}\mathbf{V}_S\mathbf{s}_S,~\textrm{if}~k=1,\\
 \sqrt{\frac{P_{R}}{d_{j,R}^{\tau_{j,R}}}} \mathbf{U}^{[k]H}_{j}\mathbf{H}_{j,R}\mathbf{V}_R\mathbf{s}_R,~\textrm{if}~k=2.
\end{cases}
\end{equation}

During the $S$-$R$ transmission, the received signal at $R$ can be written as
\begin{equation}
\mathbf{y}_{R} =  \underbrace{ \sqrt{\frac{P_{S}}{d_{R,S}^{\tau_{R,S}}}} \mathbf{U}_{R}^{H}\mathbf{H}_{R,S}\mathbf{V}_S\mathbf{{s}}_S}_{\text{desired signal}} 
+ \underbrace{ \sqrt{\frac{P_{i}}{d_{R,i}^{\tau_{R,i}}}} \sum_{i=1}^{2}\mathbf{U}_{R}^{H}\mathbf{H}_{R,i}\mathbf{V}^{[1]}_{i}\mathbf{s}_{i}}_{\text{interference from PN}} + \tilde{\mathbf{n}}_{R},
\end{equation}
where $\tilde{\mathbf{n}}_{R} = \mathbf{U}_{R}^{H}\mathbf{n}_{R}$ is the effective noise after receive beamforming at the relay. 

Then, $R$ decodes the desired signal $\mathbf{s}_S$ and forwards the information signal to $D$. Hence, $D$ receives the following signal
\begin{align}
\mathbf{y}_{D} = \sqrt{\frac{P_{R}}{d_{D,R}^{\tau_{D,R}}}} \mathbf{H}_{D,R}\mathbf{V}_{R}\mathbf{s}_R+\mathbf{n}_{D},
\end{align}
where $\mathbf{n}_{D}$ is the AWGN noise vector, with $\mathbb{E}\{\mathbf{{n}}_{D}\mathbf{{n}}^H_{D}\}=\sigma^2_{{D}}\mathbf{I}$.

The interference is assumed to be completely eliminated if the following conditions are satisfied at  $R_j$ as \cite{galym3,galym4}
\begin{subequations}
	\label{Rj condition}
	\begin{align} 
	&\mathbf{U}^{[k]H}_{j}\mathbf{H}^{[k]}_{j,i}\mathbf{V}^{[k]}_{i} = \mathbf{0},~\forall i,j\in\{1,2\},~\forall i\not=j,\\
	&\mathbf{U}^{[k]H}_{j} \mathbf{J}^{[k]} = \mathbf{0}, ~\text{where}~ \mathbf{J}^{[k]} = \begin{cases}\mathbf{H}_{j,S} \mathbf{V}_{S},~\text{if}~k = 1,\\
	\mathbf{H}_{j,R} \mathbf{V}_{R},~\text{if}~k = 2,
	\end{cases}\\
	&\textrm{rank}\left(\mathbf{U}_{j}^{[k]H}\mathbf{H}^{[k]}_{j,j}\mathbf{V}^{[k]}_{j}\right) = f_j,~\forall j\in\{1,2\}, 
	\end{align}
\end{subequations}
and at $R$ as
\begin{subequations}
	\label{R condition}
	\begin{align}
	&\mathbf{U}_{R}^{H}\mathbf{H}_{R,i}\mathbf{V}^{[1]}_{i} = \mathbf{0},~\forall i\in\{1,2\},\\
	&\textrm{rank}\left(\mathbf{U}_{R}^{H}\mathbf{H}_{R,S}\mathbf{V}_{S}\right) = f_S.    
	\end{align}
\end{subequations}

\subsection{Beamforming Design}
\label{sec:Transmit Beamforming Design}
The existing interfering signals need to be orthogonalized to ${\mathbf{U}_{j}^{[1]}}$ and ${\mathbf{U}_{j}^{[2]}}$ at $R_j$ during two TPs to satisfy conditions \eqref{Rj condition}. Then, the interference at $R$ needs to be orthogonalized to $\mathbf{U}_{R}$, as $R$ receives the interference only during the $S$-$R$ transmission. To decode the desired signal from the received signal, the interference space should be linearly independent from the desired signal space. Thus, the precoding matrices need to be designed in such a way that all interference in each receiver span to each others. Thereby, the interference at $R_1$, $R_2$ and $R$, in the first TP, can be spanned as $span(\mathbf{H}_{1,2}^{[1]}\mathbf{V}_{2}^{[1]}) = span(\mathbf{H}_{1,S}\mathbf{V}_{S})$, $span(\mathbf{H}_{2,1}^{[1]}\mathbf{V}_{1}^{[1]})=span(\mathbf{H}_{2,S}\mathbf{V}_{S})$ and $span(\mathbf{H}_{R,1}\mathbf{V}_{1}^{[1]}) = span(\mathbf{H}_{R,2}\mathbf{V}_{2}^{[1]})$, where $span(\mathbf{A})$ denotes the vector space spanned by the column vectors of $\mathbf{A}$. From these definitions, precoding matrices $\mathbf{V}_{1}^{[1]}$, $\mathbf{V}_{2}^{[1]}$ and $\mathbf{V}_{S}$ can be derived as \cite{Sung}
\begin{subequations}\label{V matirces_1}
	\begin{align}
	&\mathbf{V}_{2}^{[1]}=(\mathbf{H}_{R,2})^{-1}\mathbf{H}_{R,1}\mathbf{V}_{1}^{[1]},\\
	&\mathbf{V}_{S}=(\mathbf{H}_{2,S})^{-1}\mathbf{H}_{2,1}^{[1]}\mathbf{V}_{1}^{[1]},	
	\end{align}
\end{subequations}
where $\mathbf{V}_{1}^{[1]}=eig\left(\mathbf{A}\right)$ and $\mathbf{A}=(\mathbf{H}_{R,1})^{-1}\mathbf{H}_{R,2}(\mathbf{H}_{1,2}^{[1]})^{-1}\mathbf{H}_{1,S}(\mathbf{H}_{2,S})^{-1}\mathbf{H}_{2,1}^{[1]}$; $eig\left(\mathbf{A}\right)$ are the eigenvectors of $\mathbf{A}$. Now, we derive the corresponding receive BF matrices as 
\begin{subequations}
	\label{U matirces_1}
	\begin{align}
	&\mathbf{U}_{j}^{[k]} = null\left( [\mathbf{H}_{j,i}^{[k]}\mathbf{V}_{i}^{[k]}]^H\right),~j\not=i,\\
	&\mathbf{U}_{R} = null\left( [\mathbf{H}_{R,1}\mathbf{V}_{1}^{[1]}]^H\right).	
	\end{align}
\end{subequations}

During the $R$-$D$ transmission, $S$ stays silent while $R$ establishes its own communication. Therefore, it is clear that the BF matrices design follows the same steps given by \eqref{V matirces_1}--\eqref{U matirces_1}, and hence they are omitted for the sake of brevity.

\subsection{Imperfect CSI}
The  model for imperfect CSI can be written as \cite{galym2,Aquilina}
\begin{align}
\hat{\mathbf{H}}=\mathbf{H}+\mathbf{E},
\end{align}     
where $\hat{\mathbf{H}}$ is the observed mismatched channel, $\mathbf{H}\sim \mathcal{CN}(0,\mathbf{I})$ represents the real channel matrix and $\mathbf{E}$ is the error matrix which represents an inaccuracy degree in the estimated CSI. It is also assumed that $\mathbf{E}$ is independent of $\mathbf{H}$. Considering the nominal signal-to-noise ratio (SNR), $\theta$, $\mathbf{E}$ is described as 
\begin{align}
\mathbf{E}\sim \mathcal{CN}(0,\lambda\mathbf{I}) {\text{~with~}} \lambda=\psi{\theta}^{-\kappa},
\end{align} 
where $\lambda$ is an error variance, $\kappa\geq 0$ and $\psi>0$ determine various CSI scenarios. Finally, the real channel matrix, conditioning on ${\hat{\mathbf{H}}}$, \cite{Kay}, can be described as
\begin{align}
\label{Imperfect CSI}
\mathbf{H}=\frac{1}{1+\lambda} \hat{\mathbf{H}}+\tilde{\mathbf{H}},
\end{align}
where $\tilde{\mathbf{H}}\sim\mathcal{CN}(0, \frac{\lambda}{1+\lambda} \mathbf{I})$ is independent of $\hat{\mathbf{H}}$.

\section{Time-Switching Relaying}
\label{sec:TSR}
The time used for the $S$-$D$ information transmission is given by $T$, and the time fraction devoted for EH purposes is given by $\alpha T$, with $0\leq\alpha\leq1$. The remaining time $(1-\alpha)T$ is formed by two equal time phases to support the $S$-$R$ and $R$-$D$ transmissions \cite{Nasir}. At the same time, the PN adopts its one-hop transmission policy according to the SN time frame architecture as follows. The first data transmission occurs during the $(1+\alpha)T/2$ time because within this TP the network transmission is performed by the primary transmitters and the source node only. The remaining time $(1-\alpha)T/2$ is dedicated for the second PN transmission when the source node is replaced by the relay in the network transmission.

Hence, the received signal at $R$ during the EH phase can be written as
\begin{align}
\mathbf{y}_{R} = \sqrt{\frac{P_S}{d^{\tau_{R,S}}_{R,S}}}\mathbf{H}_{R,S}\mathbf{V}_{S}\mathbf{s}_S+\sum_{i=1}^{2}\sqrt{\frac{P_i}{d^{\tau_{R,i}}_{R,i}}}\mathbf{H}_{R,i}\mathbf{V}^{[1]}_{i}\mathbf{s}_i + \mathbf{n}_{R}.
\end{align}

Now, neglecting the power harvested from the noise, the harvested energy at $R$ is derived as \cite{Nasir}
\begin{equation}
\label{eh_tsr}
E_H^{TSR} = \eta \alpha T\left( \frac{P_S}{d^{\tau_{R,S}}_{R,S}}\left| \left|\mathbf{H}_{R,S}\mathbf{V}_S\right| \right| ^2 + \sum_{i=1}^{2}\frac{P_i}{d^{\tau_{R,i}}_{R,i}}\left| \left| \mathbf{H}_{R,i}\mathbf{V}^{[1]}_i\right| \right|^2\right),
\end{equation}
where $\parallel\cdot\parallel$ denotes the Euclidean norm.

The relay transmit power relates to the harvested energy as $P^{TSR}_R = E_H^{TSR} / ((1 - \alpha)T/2)$ and can be further rewritten as
\begin{equation}
P^{TSR}_R =\frac{2\alpha\eta}{1-\alpha} \left( \frac{P_S}{d^{\tau_{R,S}}_{R,S}}\left| \left|\mathbf{H}_{R,S}\mathbf{V}_S\right| \right| ^2 +\sum_{i=1}^{2}\frac{P_i}{d^{\tau_{R,i}}_{R,i}}\left| \left| \mathbf{H}_{R,i}\mathbf{V}^{[1]}_i\right| \right|^2\right),
\end{equation}
where $\eta$ ($0<\eta<1$) is the EH conversion efficiency. 

During the $S$-$R$ information transmission, taking into account imperfect CSI given by \eqref{Imperfect CSI} and receive BF matrices, the information signal at the relay can be written as
\begin{align}
\label{y_it_r}
\mathbf{y}_R^{IT} = \sqrt{\frac{P_S}{d^{\tau_{R,S}}_{R,S}}}{\mathbf{U}_R^H}\left( \frac{1}{1+\lambda}{\hat{\mathbf{H}}}_{R,S}+{\tilde{\mathbf{H}}}_{R,S}\right) \mathbf{V}_{S}\mathbf{s}_S + \sum_{i=1}^{2}\sqrt{\frac{P_i}{d^{\tau_{R,i}}_{R,i}}}{\mathbf{U}_R^H}\left( \frac{1}{1+\lambda}{\hat{\mathbf{H}}}_{R,i}+{\tilde{\mathbf{H}}}_{R,i}\right)\mathbf{V}^{[1]}_{i}\mathbf{s}_i + {\tilde{\mathbf{n}}}_{R}.
\end{align}
After some manipulation, the corresponding signal-to-interference-noise ratio (SINR) at $R$ can be expressed as 
\begin{align}
\label{gamma_r_ts}
\gamma_R = \frac{\frac{P_S}{d^{\tau_{R,S}}_{R,S} (1+\lambda)^2}|| \mathbf{U}_R^H\hat{\mathbf{H}}_{R,S}\mathbf{V}_S||^2}{\frac{P_S}{d^{\tau_{R,S}}_{R,S}} || \mathbf{U}_R^H\tilde{\mathbf{H}}_{R,S}\mathbf{V}_S||^2 + I_{PN}
 + \sigma^2_{\tilde{n}_R}},
\end{align}
where $I_{PN} = \frac{P_i}{d^{\tau_{R,i}}_{R,i}} \sum_{i=1}^{2} || \mathbf{U}_R^H\tilde{\mathbf{H}}_{R,i}\mathbf{V}^{[1]}_i||^2$ indicates the interference from the PN, and $\sigma^2_{\tilde{n}_R}$ stands for the noise power.  

Since it is assumed that the PN does not interfere the destination node, the signal received at $D$ can be written as
\begin{align}
\label{y_d}
\mathbf{y}_{D}=&\sqrt{\frac{P^{TSR}_R}{d^{\tau_{D,R}}_{D,R}}}\left( \frac{1}{1+\lambda}\hat{\mathbf{H}}_{D,R}+{\tilde{\mathbf{H}}}_{D,R}\right) \mathbf{V}_{R}\mathbf{{s}}_R+\mathbf{n}_{D}.
\end{align}  
Then, the respective received SINR at $D$ is calculated as
\begin{align}
\label{gamma_d_ts}
\gamma_D = \frac{\frac{P^{TSR}_R}{d^{\tau_{D,R}}_{D,R} (1+\lambda)^2 } || {\hat{\mathbf{H}}}_{D,R}\mathbf{V}_R||^2}{ \frac{P^{TSR}_R}{d^{\tau_{D,R}}_{D,R}} || \tilde{\mathbf{H}}_{D,R}\mathbf{V}_R||^2+ \sigma^2_{D} } ,
\end{align}
where $\sigma^2_{D}$ is the noise power. Finally, the received SINR for the primary receivers $R_j$ is derived as  
\begin{align}
	\label{gamma_j}
	\gamma^{[k]}_j=\frac{\frac{P_j}{d^{\tau_{j,j}}_{j,j} (1+\lambda)^2}|| {\mathbf{U}^{[k]H}_j}\hat{\mathbf{H}}^{[k]}_{j,j}\mathbf{V}^{[k]}_j||^2}{ B^{[k]} + F^{[k]}+ {\sigma_{\tilde{n}_j}^2}^{[k]}},
\end{align}
where $B^{[k]}= \frac{P_j}{d^{\tau_{j,j}}_{j,j}} || {\mathbf{U}^{[k]H}_j}\tilde{\mathbf{H}}^{[k]}_{j,j}\mathbf{V}^{[k]}_j||^2 + 
\frac{P_i}{d^{\tau_{j,i}}_{j,i}} || {\mathbf{U}^{[k]H}_j}\tilde{\mathbf{H}}^{[k]}_{j,i}\mathbf{V}^{[k]}_i||^2_{i\not=j}$ is the intra-network interference of the PN due to the CSI mismatch while the inter-network interference from the SN is expressed by
\begin{equation}
F^{[k]} = \begin{cases}
{\frac{P_S}{d^{\tau_{j,S}}_{j,S}} || {\mathbf{U}^{[k]H}_j}\tilde{\mathbf{H}}_{j,S}\mathbf{V}_S||^2},~\textrm{if}~k=1,\\
{\frac{P^{TSR}_R}{d^{\tau_{j,R}}_{j,R}} || {\mathbf{U}^{[k]H}_j}\tilde{\mathbf{H}}_{j,R}\mathbf{V}_R|| ^2},~\textrm{if}~k=2.
\end{cases}
\end{equation}

\section{Power-Splitting Relaying}
\label{sec:PSR}
The PSR protocol exploits the time $T$ divided into two equal parts to support the $S$-$R$ and $R$-$D$ information transmissions \cite{GN}. In the first TP, $R$ utilizes a portion of the received signal power for EH purposes, $\rho$, while the rest of the power, $(1 - \rho)$, is dedicated for the $S$-$R$ data transmission \cite{Arzykulov1}.
Thus, the relay utilizes the energy harvested from the received signal given by 
\begin{equation}
\mathbf{y}_R^{EH}=\sqrt{\frac{\rho P_S}{d^{\tau_{R,S}}_{R,S}}}\mathbf{H}_{R,S}\mathbf{V}_{S}\mathbf{s}_S + \sum_{i=1}^{2}\sqrt{\frac{\rho P_i}{d^{\tau_{R,i}}_{R,i}}}\mathbf{H}_{R,i}\mathbf{V}^{[1]}_{i}\mathbf{s}_i+\sqrt{\rho}\mathbf{n}_{R}.
\end{equation}

By assuming that the received signal $\mathbf{y}_R^{EH}$ at $R$ is used only for WPT and the noise power is neglected, the instantaneous harvested energy can be expressed as \cite{Nasir}
\begin{equation}
\label{E_h}
E_H^{PSR} = \frac{\eta\rho T}{2}\left( \frac{P_S}{d^{\tau_{R,S}}_{R,S}}\left| \left|\mathbf{H}_{R,S}\mathbf{V}_S\right| \right| ^2 +\sum_{i=1}^{2}\frac{P_i}{d^{\tau_{R,i}}_{R,i}}\left| \left| \mathbf{H}_{R,i}\mathbf{V}^{[1]}_i\right| \right|^2 \right), 
\end{equation}
where $0<\rho<1$ is the signal power portion dedicated for EH purposes. The relay transmit power as a function of the harvested energy is given by $P^{PSR}_R=2 E_H^{PSR}/T$, and it can be then written as \cite{Nasir}
\begin{equation}
P^{PSR}_R=\eta\rho\left( \frac{P_S}{d^{\tau_{R,S}}_{R,S}}\left| \left|\mathbf{H}_{R,S}\mathbf{V}_S\right| \right| ^2 +\sum_{i=1}^{2}\frac{P_i}{d^{\tau_{R,i}}_{R,i}}\left| \left| \mathbf{H}_{R,i}\mathbf{V}^{[1]}_i\right| \right|^2 \right).
\end{equation}

Using \eqref{Imperfect CSI}, the same signal, but with the power portion of $(1-\rho)$ and with the receive BF matrix $\mathbf{U}_R$ applied, can be received at the information decoder terminal as shown in \eqref{y_IT} at the top of the next page.
\begin{figure*}[!t]
	\small
	\begin{align}
	\label{y_IT}
	\mathbf{y}^{IT}_R =& \sqrt{1-\rho} \left( \sqrt{\frac{P_S}{d^{\tau_{R,S}}_{R,S}}} \mathbf{U}^H_R\left( \frac{1}{1+\lambda}\hat{\mathbf{H}}_{R,S} + \tilde{\mathbf{H}}_{R,S}\right) \mathbf{V}_{S}\mathbf{s}_S + \sum_{i=1}^{2} \sqrt{\frac{P_i}{d^{\tau_{R,i}}_{R,i}}} \mathbf{U}^H_R \left( \frac{1}{1+\lambda}\hat{\mathbf{H}}_{R,i}+\tilde{\mathbf{H}}_{R,i} \right) \mathbf{V}^{[1]}_{i}{\mathbf{s}_i} + \tilde{\mathbf{n}}_{R} \right).
	\end{align}
	\hrulefill
\end{figure*}
Now, by using \eqref{y_IT}, the SINR of the $S$-$R$ link can be derived as
\begin{align}
\label{gamma_r_ps}
\gamma_R=\frac{\frac{P_S(1-\rho)}{d^{\tau_{R,S}}_{R,S} (1+\lambda)^2}|| \mathbf{U}_R^H\hat{\mathbf{H}}_{R,S}\mathbf{V}_S|| ^2}{\frac{P_S (1-\rho)}{d^{\tau_{R,S}}_{R,S}} || \mathbf{U}_R^H\tilde{\mathbf{H}}_{R,S}\mathbf{V}_S||^2 + I_{PN} + \sigma^2_{\tilde{n}_R}},
\end{align}
where $I_{PN} = \frac{P_i (1-\rho)}{d^{\tau_{R,i}}_{R,i}} \sum_{i=1}^{2} || \mathbf{U}_R^H\tilde{\mathbf{H}}_{R,i}\mathbf{V}^{[1]}_i||^2$ denotes the interference from the PN.

Then, the corresponding SINR at $D$ is calculated as
\begin{align}
\label{gamma_d_ps}
\gamma_D = \frac{\frac{P^{PSR}_R}{d^{\tau_{D,R}}_{D,R} (1+\lambda)^2 } || {\hat{\mathbf{H}}}_{D,R}\mathbf{V}_R||^2}{ \frac{P^{PSR}_R}{d^{\tau_{D,R}}_{D,R}} || \tilde{\mathbf{H}}_{D,R}\mathbf{V}_R||^2+ \sigma^2_{D} } ,
\end{align}
where $\sigma^2_{D}$ is the noise power. 

Finally, the SINR of the primary users can be calculated as 
\begin{align}
	\label{gamma_j_ps}
	\gamma^{[k]}_j=\frac{\frac{P_j}{d^{\tau_{j,j}}_{j,j} (1+\lambda)^2}|| {\mathbf{U}^{[k]H}_j}\hat{\mathbf{H}}^{[k]}_{j,j}\mathbf{V}^{[k]}_j||^2}{B^{[k]} + F^{[k]} + {\sigma_{\tilde{n}_{j}}^2}^{[k]}},
\end{align}
where $B^{[k]} = \frac{P_j}{d^{\tau_{j,j}}_{j,j}} || {\mathbf{U}^{[k]H}_j}\tilde{\mathbf{H}}^{[k]}_{j,j}\mathbf{V}^{[k]}_j ||^2 + 
\frac{P_i}{d^{\tau_{j,i}}_{j,i}} || {\mathbf{U}^{[k]H}_j}\tilde{\mathbf{H}}^{[k]}_{j,i}\mathbf{V}^{[k]}_i ||^2_{i\not=j}$ is the intra-network interference of the PN due to the CSI mismatch while the inter-network interference from the SN is expressed by
\begin{equation}
F^{[k]} = \begin{cases}
{\frac{P_S}{d^{\tau_{j,S}}_{j,S}} || {\mathbf{U}^{[k]H}_j}\tilde{\mathbf{H}}_{j,S}\mathbf{V}_S|| ^2},~\textrm{if}~k=1,\\
{\frac{P^{PSR}_R}{d^{\tau_{j,R}}_{j,R}} || {\mathbf{U}^{[k]H}_j}\tilde{\mathbf{H}}_{j,R}\mathbf{V}_R||^2},~\textrm{if}~k=2.
\end{cases}
\end{equation}	

\section{Ergodic Capacity}
\label{sec:capacity}
For the PN receivers, the general derivation of the instantaneous capacity can be written as \cite{Tang}
\begin{align}
\mathcal{C}_j = \log_2(1+\gamma_j),
\end{align}
where $\gamma_j$ is the instantaneous SINR at $R_j$. Regarding the SN operating in the DF mode, the end-to-end capacity at $D$ is related to the weakest link of the $S$-$R$ and $R$-$D$ transmissions and can be written as \cite{Arzykulov}
\begin{align}
\label{capacity}
\mathcal{C}_D = \frac{1}{2} \log_2\left(1+\min\left(\gamma_R, \gamma_D\right)\right),
\end{align}
where $\gamma_R$ and $\gamma_D$ denote the instantaneous SINR at $R$ and $D$, respectively.

Now, we derive an expression for the capacity of the
TSR-based system. Using \eqref{gamma_r_ts}, \eqref{gamma_d_ts} and \eqref{capacity}, the capacity of the destination node can be written as
\begin{align}
\label{C_d_tsr}
\mathcal{C}_D = \frac{1-\alpha}{2}\log_2\left(1 + \min\left(\gamma_R,\gamma_D\right)\right).
\end{align}

Regarding the PN, the capacity of $R_j$ can be written as
\begin{equation}
\mathcal{C}_j = \sum_{k=1}^{2} E^{[k]} \log_2(1+\gamma_j^{[k]}),~j\in\{1,2\},
\end{equation}
where
${E}^{[k]} = \begin{cases}
{\frac{1+\alpha}{2}},~\textrm{when}~k=1,\\
{\frac{1-\alpha}{2}},~\textrm{otherwise},
\end{cases}$ with $k\in\{1,2\}$.

Using \eqref{gamma_r_ps}, \eqref{gamma_d_ps} and \eqref{capacity}, the capacity of the destination node for the PSR-based system can be written as
\begin{align}
\label{C_d_psr}
\mathcal{C}_D = \frac{1}{2}\log_2\left(1 + \min\left(\gamma_R,\gamma_D\right)\right).
\end{align}

Regarding the PN, the capacity of $R_j$ can be written using \eqref{gamma_j_ps} as
\begin{equation}
\mathcal{C}_j = \frac{1}{2} \sum_{k=1}^{2} \log_2(1+\gamma_j^{[k]}),~j\in\{1,2\}.
\end{equation}

We evaluate the capacity of the optimized TSR and PSR-based systems as a function of SNR. To begin with, we find the optimal values of $\alpha$ and $\rho$ given by $\alpha^*$ and $\rho^*$ with $\eta = 0.8$ for the corresponding TSR and PSR protocols by solving the following $d \mathcal{C}_D / d\alpha= 0$ and $d \mathcal{C}_D / d\rho= 0$.

\section{Simulation Results}
\label{sec:numerical}
In this section, we present numerical examples for the capacity expressions derived above. The adopted system parameters are as follows: the channel distances and path loss exponents are identical and given by $d = 3$ m and $\tau = 2.7$, respectively; the EH conversion efficiency $\eta = 0.8$; the fixed transmit powers are assumed to be equal ($P_1 = P_2 = P_S$). The calculated optimal EH time fraction $\alpha$ and power-splitting factor $\rho$ are taken as 0.19 and 0.75, accordingly. Finally, the values of $(\kappa,\psi)$ such as  $(1.5,15),~(1,10),~(0,0.001)$ are considered to describe various CSI mismatch scenarios.

\begin{figure*}[!h]
	\centering
	\subfloat[Perfect CSI vs. CSI mismatch $(\kappa=0,~\psi=0.001)$.]{
		\label{subfig:cap_snr_perf}
		\includegraphics[width=0.4\columnwidth]{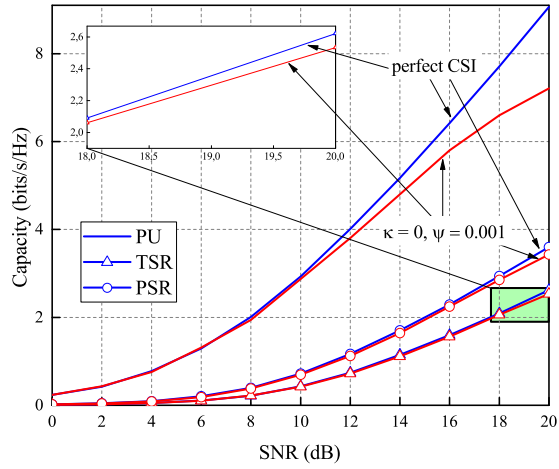}}~~~~
	\subfloat[CSI mismatch: $(\kappa=1.5,~\psi=15)$ vs. $(\kappa=1,~\psi=10)$.]{
		\label{subfig:Nakagami_d}
		\includegraphics[width=0.4\columnwidth]{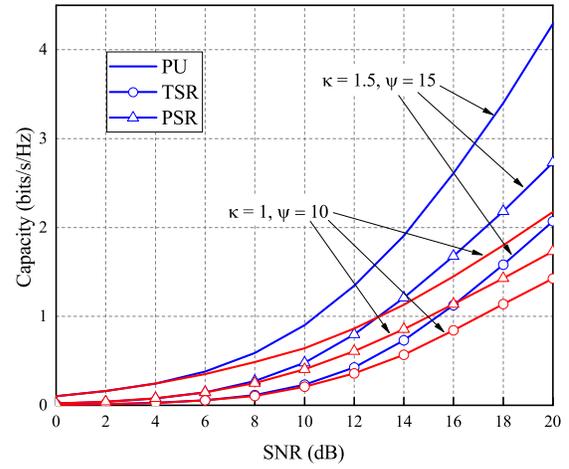}}
	\caption{Capacity vs. SNR of the primary user and the destination node operating in the TSR and PSR protocols under different CSI scenarios.}
	\label{results1}
\end{figure*} 
Fig. \ref{results1} presents an insight into how the CSI mismatch affects the obtainable capacity of the PUs and the SU of the considered system model. When $\kappa=0$, it can be noticed that $\psi$ has a significant impact on the system capacity. At $\psi=0.001$ and SNR of 20 dB, the capacity loss of the PUs equals 1.85 bit/s/Hz while the capacity losses of the destination node for the TSR and PSR protocols are given by 0.09 and 0.17 bit/s/Hz, respectively. Hence, it is obvious that the CSI mismatch severely degrades the performance of the PUs which can be explained by the higher level of the interference due to the number of the transmitting nodes. In general, the capacity of the destination node of the PSR-based system always outperforms the capacity of the TSR protocol even the former relaying method experiences worse performance degradation than the latter one. When $\kappa\not=0$, it can be noticed that the capacity demonstrates a poor performance in the low SNR region. On the other side, the dependence on the SNR results in the capacity growth when SNR increases. Being $\kappa$ large makes the curve slope sharper.

\begin{figure}[!t]
	\centering
	\includegraphics[width=0.4\columnwidth]{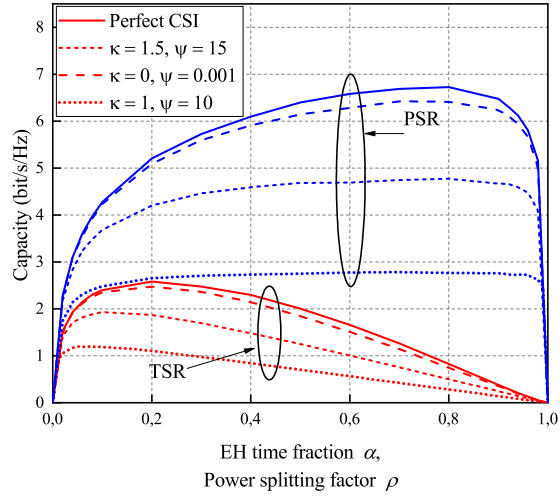}
	\caption{Capacity vs. the EH time fraction and power splitting factor for the TSR and PSR at 20 dB, respectively.}
	\label{results2}
\end{figure}

Fig. \ref{results2} illustrates some simulated results for the capacities as a function of $\alpha$ and $\rho$ for various CSI acquisition scenarios. The results for the TSR and PSR systems are obtained from \eqref{C_d_tsr} and \eqref{C_d_psr}, respectively. For the case of the PSR-based system, no sufficient portion of power allocated for EH will consequently result in poor capacity. At the other extreme, being $\rho$ too large which results in too much unnecessarily harvested energy at the cost of the power level of the received signal leading to poor capacity. Similarly, this justification applies to $\alpha$ in the TSR-based system. 

\section{Conclusion}
\label{sec:Conclusion}
In this paper, we analyzed the capacity of different wireless powered IA-based DF CRN considering the EH protocols, namely, TSR and PSR. The three special scenarios of the imperfect CSI given by $(1.5,15),~(1,10),~(0,0.001)$ were studied to analyse the impact of the CSI quality on the system capacity. The presented results demonstrated that an optimal selection of the power splitting factor in the PSR protocol and the EH time factor in the TSR protocol were found to be important in achieving the best performance. Finally, the optimized PSR-based system was shown to have the best performance over the optimized TSR system. 

\ifCLASSOPTIONcaptionsoff
  \newpage
\fi

%





\end{document}